# Continuous-wave mirrorless lasing at 2.21 µm in sodium vapors

ALEXANDER M. AKULSHIN[1,2*], FELIPE PEDREROS BUSTOS[2], AND DMITRY BUDKER[2,3]

[1]*Centre for Quantum and Atom Optical Science,*
*Swinburne University of Technology, PO Box 218, Melbourne 3122, Australia*
[2] *Helmholtz Institute, Johannes Gutenberg University, D-55128 Mainz, Germany*
[3] *Department of Physics, University of California, Berkeley, CA 94720-7300, USA*
*\*Corresponding author: aakoulchine@swin.edu.au*

**We demonstrate backward-directed continuous-wave (cw) emission at 2.21 µm generated on the $4P_{3/2}$-$4S_{1/2}$ population-inverted transition in Na vapors two-photon excited with resonant laser light at 589 and 569 nm. Our study of power and atom-number-density threshold characteristics shows that lasing occurs at sub-10 mW total power of the applied laser light. The observed 6 mrad divergence is defined mainly by the aspect ratio of the gain region. We find that mirrorless lasing at 2.21 µm is magnetic field and polarization dependent that may be useful for remote magnetometry. The presented results could help determine the requirements for obtaining directional return from sodium atoms in the mesosphere.**

Thorough investigations of a variety of nonlinear effects in atomic media, such as stimulated hyper-Raman scattering, amplified spontaneous emission (ASE) and parametric wave mixing were previously conducted in atomic media with high number density (N > $10^{14}$ cm$^{-3}$) using short laser pulses with power above the $10^5$ W level [1-5]. These studies were primarily motivated as development of sources of coherent radiation for spectroscopic applications. Later, it was shown that comparable efficiency of nonlinear processes can be achieved even with orders of magnitude weaker continuous-wave (cw) laser light tuned to dipole-allowed transitions in atomic media. In this case, the generation of new optical fields occurs at discrete wavelengths associated with certain transitions in atomic media [6]. In recent years, frequency up- and down-conversion, i.e. new field generation, due to nonlinear processes in alkali vapors excited with cw low-power (less than 100 mW), resonant laser light received considerable attention [7-11] due to numerous promising applications, including quantum memory [12] and correlated-photon generation [13]. It was shown that generation of directional laser-like radiation can be attributed to four-wave mixing (FWM) and amplified spontaneous emission (ASE) [9, 14]. However, the interplay of parametric and nonparametric nonlinear processes responsible for the new-field generation is not entirely understood. The cw low-power approach is attractive for studying nonlinear processes with sub-Doppler spectral resolution.

Investigations of nonlinear mixing and directional light generation using cw lasers have been carried out almost exclusively using Rb atoms. Here, motivated by possible applications in remote detection [15-17] and, particularly, by the possibility of generating backward-directed emission from the mesospheric sodium layer [18, 19], we report on the experimental observation of the cw mirrorless lasing on the $4P_{3/2}$-$4S_{1/2}$ transition in Na vapors (Fig. 1) and the study of its spectral and spatial characteristics.

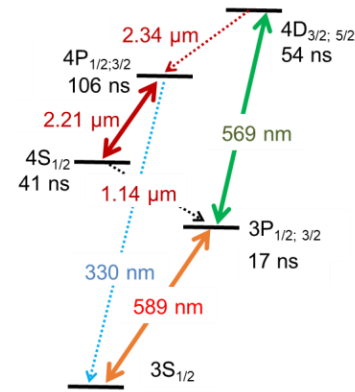

Fig. 1. Energy levels and their lifetimes [20] of Na atom involved in two-photon excitation to the 4D level by laser fields at 589 and 569 nm.

Population inversion on the $4P_{3/2}$-$4S_{1/2}$ transitions can be prepared by exciting Na atoms directly from the ground state to the $4P_{3/2}$ level using resonant UV radiation at 330 nm as was demonstrated in [21]. However, in our previous experiment with two-photon excited Rb vapors, generation of forward-directed emission at 2.8 µm on the $6P_{3/2}$-$6S_{1/2}$ transition that is an analogue to the $4P_{3/2}$-$4S_{1/2}$ transition in Na, revealed itself through the generation of another directional light at 1.37 μm via the FWM process [9]. Therefore, we commence our study of the directional emission in Na vapors with the bi-chromatic excitation, similar to the one used in our experiments with Rb [9-11].

In our experiment, we use two commercial cw single-mode dye lasers (Coherent 699-21 and 899-21) with short-term linewidth of 5 MHz and 15 MHz, respectively. The lasers are locked to tuneable Fabry–Perot cavities with less than the 5 MHz/min long-term drift. Absolute optical frequencies of the lasers are monitored with a wavemeter (High Finesse WS6-600).

The sodium atoms are contained in a quartz buffer-gas-free cell of 3 cm diameter and 10 cm length. A carefully designed heater provides a temperature gradient along the cell that prevents Na atoms from condensing on the windows. The cell temperature is limited to 185 °C, minimizing chemical reaction between hot sodium vapors and the cell.

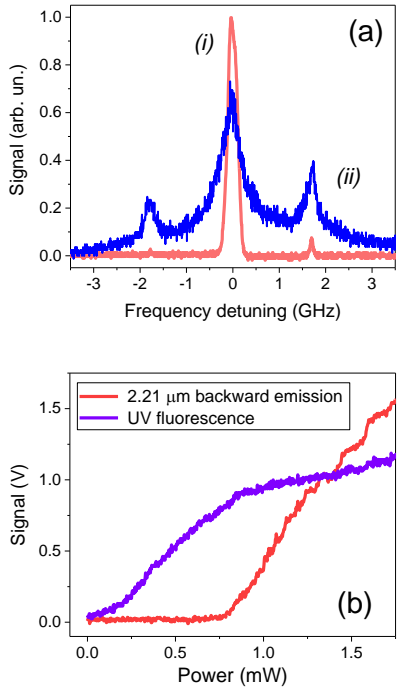

Fig. 2. (a) Backward-directed IR emission at 2.21 μm (*i*) and transverse-detected UV fluorescence (*ii*) from Na vapors two-photon excited by co-propagating beams at 589 and 569 nm as a function of the frequency detuning of the 589 nm laser from the $3S_{1/2}(F=2)$ - $3P_{3/2}$ transition. The frequency of the 569 nm laser is tuned to the $3P_{3/2}$-$4D_{5/2}$ transition. The applied laser power at 589 and 569 nm is 20 and 25 mW, respectively. The atom number density in the cell is $N \approx 3.0\times10^{11}$ cm$^{-3}$. (b) Emission at 2.21 μm and UV fluorescence as a function of laser power at 569 nm, while laser power at 589 nm is 20 mW and $N \approx 3.5\times10^{11}$ cm$^{-3}$. The lasers at 589 and 569 nm are tuned to the $3S_{1/2}(F=2)$-$3P_{3/2}$ and $3P_{3/2}(F=2)$-$4D_{5/2}$ transitions, respectively.

Fluorescence is a commonly used indicator of the number of excited atoms. The fluorescence at 330 nm emitted by cascade-decaying Na atoms from the excited 4D level is detected with a photomultiplier tube (PMT). Spectroscopic signals are recorded directly or after lock-in amplification. In the latter case, the applied laser light is mechanically chopped at 360 Hz.

First, we excite Na vapors with counter-propagating laser beams at 589 and 569 nm, as such geometry provides a higher number of excited atoms. At certain atom number density N and appropriate power of the laser fields tuned to the $3S_{1/2}(F=2)$-$4D_{5/2}$ two-photon transition, directional emission appears on the $4P_{3/2}$-$4S_{1/2}$ population-inverted transition. However, as the co-propagating geometry of excitation might provide, under certain conditions, backward-directed laser-like emission that is crucial for potential applications in remote detection [15-17], we focus on studying stimulated emission generated in Na vapors excited by the laser beams propagating collinearly. For this purpose, radiation from the two lasers are combined on a non-polarizing beam splitter to form two co-propagating bi-chromatic beams. One of them is loosely focused inside the cell so that the minimal cross section of the 589 and 569 nm components are approximately 250 and 350 μm, respectively. Thus, the aspect ratio of the two-photon interaction region is defined by the 589 nm component.

As shown in [22], velocity-selective ground-state hyperfine repumping can significantly enhance the efficiency of the parametric four-wave mixing (FWM) process. In our experiment, instead of employing an additional laser to produce the required repumping field, the laser light at 589 nm is frequency modulated at 1.713 GHz with an electro-optical modulator (EOM). In this case both the carrier and the high-frequency sideband simultaneously interact with atoms from the resonant velocity group on the cycling $3S_{1/2}(F=2)$-$3P_{3/2}(F'=3)$ and open $3S_{1/2}(F=1)$- $3P_{3/2}(F'=2)$ transitions that increases the number of atoms on the $3S_{1/2}(F=2)$ level in the resonant group.

Figure 2a shows that spectral profiles of lateral UV fluorescence and the 2.21 μm emission detected 25 cm away from the cell in the backward direction are very different. The UV fluorescence is observed in a wide range of frequency detuning of the 589 nm laser from the $3S_{1/2}(F=2)$-$3P_{3/2}(F'=3)$ transition defined by the inhomogeneous broadening of the Na-D2 absorption line. As frequency modulation is applied, the UV profile consists of three peaks spectrally separated by 1.713 GHz. The directional IR emission occurs in two spectral regions. At zero detuning, the high-frequency sideband at 589 nm interacts with atoms from resonant velocity group on the open transition that results in higher number of excited atoms on the $4D_{5/2}$ level, in higher population inversion on the $4P_{3/2}$-$4S_{1/2}$ transition and, eventually, in the generation of collimated emission at 2.21 μm along the pencil-shaped region that contains population-inverted atoms. The directional IR emission also occurs when the low-frequency sideband interacts with atoms from the resonant velocity group on the $3S_{1/2}(F=2)$-$3P_{3/2}(F'=3)$ transition, while the carrier interacts with them mainly on the open $3S_{1/2}(F=1)$-$3P_{3/2}(F'=2)$ transition. Both peaks have sub-Doppler width; however, the smaller peak is substantially narrower, its FWHM is ~80 MHz, compared to 250 MHz for the large peak. We explain this by a threshold effect.

Pronounced threshold dependence of the directional emission at 2.21 μm on the applied laser power at 569 nm is shown in Fig. 2b. At low power, $P_{589} < 0.7$ mW, lateral UV fluorescence at 330 nm displays a noticeable growth with the applied power, but the backward emission at 2.21 μm is hardly detectable. However, once the stimulated process prevails over spontaneous decay at $P_{589} > 0.75$ mW, the IR emission grows sharply. We also note that, the slope of the UV fluorescence curve is reduced above the ASE threshold because the stimulated process on the $4P_{3/2}$-$4S_{1/2}$ transition changes the balance between the decay pathways at 330 nm and 2.21 μm.

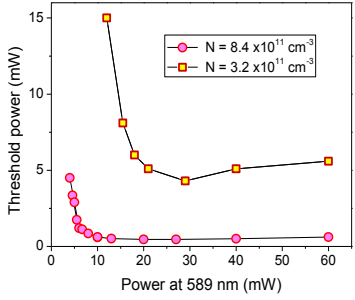

Fig. 3. The 569 nm ASE threshold power as a function of the applied laser power at 589 nm at different atom number density N in the cell.

The observed threshold behaviour of the 2.21 μm emission and its narrow spectral linewidth relative to the fluorescence profile indicate the onset of laser-like emission. We attribute this emission to the ASE process [23, 24] that occurs in the elongated interaction region defined by the overlap of the co-propagating laser beams inside the cell.

The ASE threshold is a nontrivial function of the applied laser power and atom number density N, as shown in Figure 3. For both atom number densities considered here, the threshold power at 569 nm first decreases sharply with the applied 589 nm power and after reaching the minimum rises again.

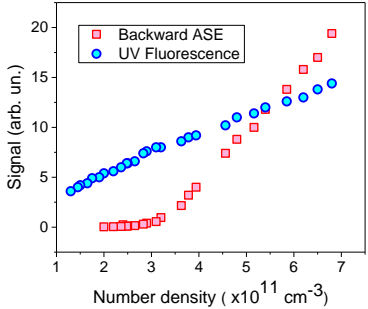

Fig. 4. Backward emission at 2.21 μm and UV fluorescence as a function of atom number density N in the cell. The applied laser powers at 589 and 569 nm is 30 mW and 12 mW, respectively, while the laser frequencies are tuned to the maximal power of mirrorless lasing.

For a given applied laser power, the backward ASE at 2.21 μm generated by the bi-chromatic laser beam at 589 and 569 nm has a threshold dependence on atom number density N in the cell, as shown in Figure 4. At $N \leq 3 \times 10^{11}$ cm$^{-3}$, the UV fluorescence signal grows steadily with atom number density N, while the IR emission at 2.21 μm is nearly isotropic and weak. However, above the threshold value ($N \approx 3.2 \times 10^{11}$ cm$^{-3}$), the gain of population inverted atoms becomes large enough to start mirrorless lasing. As a result, the IR signal grows linearly to N in the $3.5 \times 10^{11}$ cm$^{-3}$ $\leq N \leq 7 \times 10^{11}$ cm$^{-3}$ atom number-density range without noticeable saturation.

As the distribution of excited atoms among the hyperfine and magnetic sub-levels of the $4D_{5/2}$ level is important for reaching the ASE threshold, the lasing power depends on the polarization of the applied laser light and magnetic fields. We find that more efficient excitation occurs when both components of the applied laser light have the same circular polarization. In this case the applied laser light also provides the highest excitation probability for the stepwise process. With the present experimental parameters, the maximum ASE power at 2.21 μm obtained with linearly polarized bi-chromatic beam was approximately ten times weaker. Higher efficiency of the parametric FWM process for circularly polarized laser fields was also demonstrated in Rb vapors [14].

The mirrorless lasing reveals strong magnetic field dependence (Fig. 5). We find that the lasing occurs in a narrower interval of transverse magnetic fields for lower light power and that it is most sensitive to magnetic fields just above the threshold. A detailed study of the magnetic effects in directional emission from sodium vapors will be provided in a forthcoming publication. This may be applicable for remote magnetometry with mesospheric sodium, as fluorescence resonances at 589 nm with the 0.2 μT-width have been observed [17].

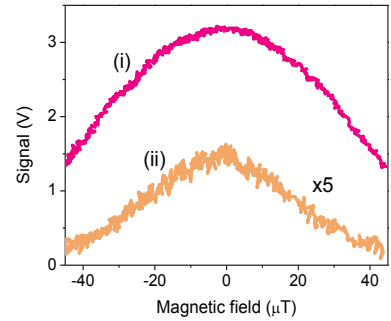

Fig. 5. Backward emission at 2.21 μm as a function of a magnetic field applied in a direction transverse to the light-propagation axis at different power of the applied laser light. Curve (*i*) is recorded with 18 mW and 60 mW at 589 and 569 nm, respectively, while curve (*ii*) is taken with 7 mW and 2.4 mW at 589 and 569 nm, respectively. Both the lasers are tuned to the maximal intensity of the backward-directed emission at $N \approx 5 \times 10^{11}$ cm$^{-3}$.

The directionality of mirrorless lasing due to the ASE process, as was point out in [25], is primarily determined by the aspect ratio of the gain region, assuming a homogeneous distribution of population-inverted atoms in a pencil-shaped region. In our case, the aspect ratio is $\varnothing / \ell \approx 2.5 \times 10^{-3}$, where $\varnothing \approx 0.025$ cm and $\ell = 10$ cm are the minimum diameter and the length of the gain region, respectively. Figure 5b demonstrates the spatial profiles of backward lasing recorded by scanning the lasing beam across the IR detector at a distance of 300 cm from the cell. A conservative estimate shows that the divergence does not exceed 6 mrad. This value is close to the diffraction limit for Gaussian beams at λ = 2.21 μm, $\theta_D \approx \lambda/\pi w_0 \approx 3$ mrad, and to the estimated aspect ratio, considering that the cross section of the gain region is not constant inside the cell. The relation between the geometry of the interaction region and gain variations in the transverse and longitudinal directions on the one hand and the ASE threshold and spatial characteristics of the mirrorless lasing on the other, deserve additional theoretical and experimental investigation.

The presented study may be useful for developing new methods of remote sensing, in particular, for enhancing the efficiency of the laser guide star (LGS) techniques [18]. The conventional LGS method for image correction in astronomy is based on the detection of nearly isotropic fluorescence from mesospheric Na atoms [18]. Despite the significant progress in recent years, there is a demand for stronger return that is needed for faster and more accurate image correction. Our approach using cw resonant excitation may turn out to be more practical compared with recently obtained backward cooperative emission from dense Na vapors using ultrashort pulses produced by an amplified femtosecond laser [19].

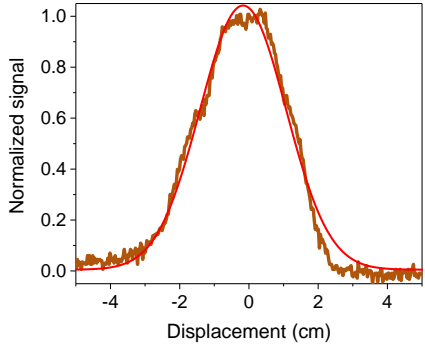

Fig. 6. Spatial profiles of backward-directed ASE emission at 2.21 μm taken by scanning the beam across the detector with the 3×3 mm$^2$ light sensitive region at 300 cm from the cell center. This corresponds to the divergence half-angle of about 6 mrad.

The maximum power of the cw backward ASE at 2.21 μm obtained at the atom number density N ≈ 8×10$^{11}$ cm$^{-3}$ is approximately 30 μW, while the applied laser power at 589 and 569 nm is 70 mW and 60 mW, respectively. The conversion efficiency of the applied laser light into backward emission is 2.5×10$^{-4}$. This is approximately 70 times higher compared to the efficiency value obtained for backward emission from more dense Na vapors (N ≈ 1.7×10$^{16}$ cm$^{-3}$) in a heat-pipe oven using ultrashort pulses [19].

Minimization of power and atom number density thresholds of the ASE-based lasing is crucially important for possible applications for remote detection. Recently it was shown that there is an optimal power ratio between the components of the bi-chromatic beam for maximizing the efficiency of new field generation by parametric FWM [26]. A detailed study of the optimal conditions minimizing the power and atom number density thresholds is the subject of the forthcoming paper.

In conclusion, we have experimentally investigated the properties of continuous-wave backward-directed 2.21 μm emission generated in Na vapors excited to the 4D$_{5/2}$ level with resonant laser light at 589 and 569 nm in the co-propagating configurations. It has been shown that lasing can occurs at sub-10 mW total power of the applied laser light.

The observed threshold atom number-density and laser power dependences as well as the estimated divergence of the backward-directed emission are consistent with the ASE mechanism.

The threshold of mirrorless lasing and the generated power are sensitive to the polarization of the laser light and to magnetic fields, the latter being is promising for remote magnetometry.

The results of our study of parametric and nonparametric nonlinear processes in sodium vapors performed with sub-Doppler spectral resolution may provide decisive information for choosing an optimal excitation scheme for achieving directional return from mesospheric sodium atoms, as it promises a dramatic enhancement of LGS signals. The current approach for generating directional backward emission for remote detection is applicable to a wide range of atomic and plasma media.

**Acknowledgments**. This work has been supported in part by the US Office of Naval Research Global under grant N62909-16-1-2113. We thank the European Southern Observatory for the loan of a dye-laser system. F.P.B. acknowledges the support from a Carl-Zeiss Foundation Doctoral Scholarship.